*2024 CRA Quadrennial Paper*

# Empowering the Future Workforce: Prioritizing Education for the AI-Accelerated Job Market

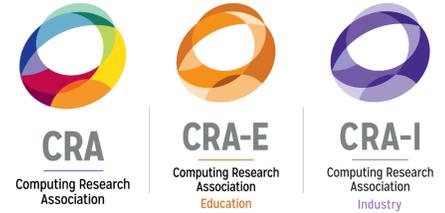

Lisa Amini (IBM Research), Henry F. Korth (Lehigh University), Nita Patel (Otis), Evan Peck (University of Colorado Boulder), Ben Zorn (Microsoft)

**AI's rapid integration into the workplace demands new approaches to workforce education and training and broader AI literacy across disciplines. Coordinated action from government, industry, and educational institutions is necessary to ensure workers can adapt to accelerating technological change.**

It is believed by some that we are entering a new age of technology, characterized by advanced, pervasive Artificial Intelligence (AI), during which the rate of workforce and economic disruption will be substantially greater than previous periods. Regardless of whether a new era has commenced, AI is increasing in capability, speeding integration into the workplace and our homes, and prevailing in both technical and non-technical contexts and occupations.

New skills and professions — many of which are not yet conceived — will arise, as will widespread job displacement. Just as the Information Age required national imperatives for computing education, similar imperatives exist for the rise of AI. In a survey of 4702 CEOs, 70 percent say AI will significantly change the way their companies create, deliver, and capture value over the next three years, and 45 percent believe their companies will no longer be viable in ten years if they continue on their current path. The 2024 AI Jobs Barometer analysis of over a half billion job ads identified rapid and significant shifts in the skills needed in both AI-specific and AI-exposed jobs. Beyond technical advancements, improving AI literacy in the workforce is critical to building trust, resisting disinformation from bad actors, and refining human-AI interaction to prevent harm. **Figure 1** illustrates a greater change in skills demanded for AI-exposed jobs is already underway.

**Figure 1:** *Source: PwC analysis of Lightcast data, ISCO-08 Occupation Codes (2-digit level). The net skill change is calculated as the difference between 2019-2023 in the total number of skills required by job occupations using the ISCO-08 4-digit occupational codes.*

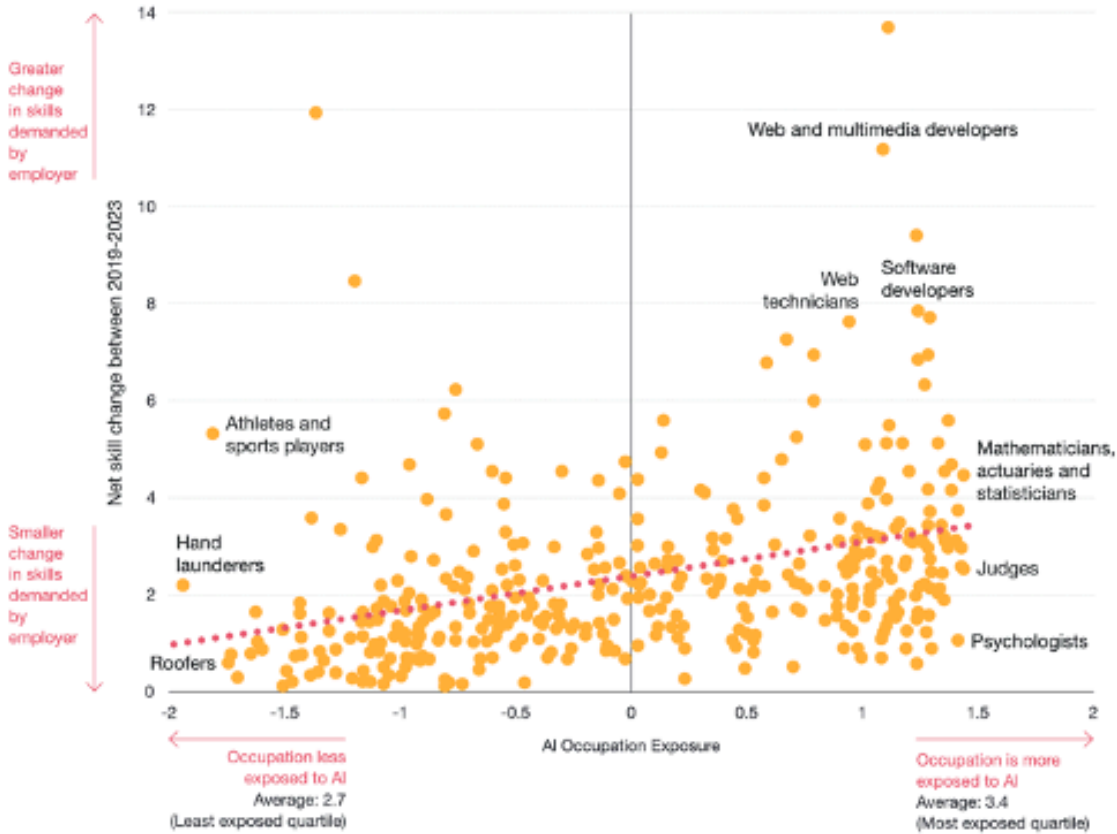

This report proposes actions that government, industry, and academia can take to better prepare our workforce for AI-accelerated shifts in skills and jobs. This includes identifying current barriers to AI education and lifelong AI literacy, as well as opportunities for more rapid skill evolution and workforce resiliency. We propose educational requirements to help build foundational AI knowledge regarding AI and advocate for the role of industry in reskilling efforts.

## Opportunities for Impactful, Broad-Based Education

Significant technological and resource barriers threaten to impede workforce preparedness for AI's impact. At the same time, AI's increasing cognitive abilities present new opportunities to reach a larger percentage of the workforce, including those without formal computing education.

## Barriers and Opportunities

- **Scope of AI Education:** Although AI is expected to disrupt the workforce broadly, AI education currently is predominantly situated within formal computing disciplines. An important insight from the Information Age — one that applies even more urgently in the era of advanced AI and robotics — is that AI education needs extend well beyond computing contexts.

- **Technological and Cost Barriers:** Leading-edge AI currently demands large and costly computational resources and technical expertise. For example, the technical creation cost of one version of one particular model, *ChatGPT*-4, has been estimated as $41 million to $78 million. Compute infrastructures for training leading-edge models employ tens of thousands of graphics processing units (GPU). While accessing AI does not require creating one's own model, it does require at minimum paying a fee to access the best models or compute infrastructure to run freely downloadable models.

- **Accelerating Rate of Change:** UBS analysts reported that ChatGPT may have broken the record for the fastest-growing app in history, reaching an estimated 123 million monthly active users less than three months after launch. In comparison, the time previous technologies took to reach 50 percent of U.S. users was 20 years for PCs, 12 years for the internet, and six years for smartphones.

- **Natural Language as a Computer Programming Language**: Previous generations of software required learning specialized skills such as software architecture, design, and engineering, and programming languages, such as COBOL, Java, and Python. Today's large language models (LLMs) enable humans to interact with LLMs through natural language, leading to the observation that natural language is becoming a significant computer programming language in its own right. While natural language interfaces to AI are currently most pervasive, interfaces based on other modalities — such as visual, motion, and audio — are also on the rise.

- **The Cost of AI Decision-Making**: While generative AI is widely seen as transformational, it is not without issues. Integrating AI into critical decision-making environments — including healthcare, autonomous vehicles, policing, housing allocation, and insurance — carries an inherent risk of causing severe harm, particularly to marginalized communities. AI has already demonstrated the capacity for disinformation, such as an AI-generated article falsely claimed: "In a bizarre turn of events, NBA star Klay Thompson has been accused of vandalizing multiple houses in Sacramento," after misinterpreting social media references to Thompson "shooting bricks" (a slang term for repeatedly missing shots).

- **Security, Privacy, and Safety Risks**: Protecting private or sensitive data is another challenge in AI adoption. User inputs, including prompts, may be incorporated into training datasets, leading to the risk of data leakage. Security threats, such as intentional poisoning of training datasets or adversarial attacks on AI models, further complicate AI adoption. AI safety concerns include ensuring compliance with legal and ethical standards, avoiding bias, and preventing misuse.

While these issues and opportunities extend into K-12 education, that discussion is beyond the scope of this paper. Additionally, broader concerns around data literacy, critical thinking, and adaptability to technological change suggest additional areas of focus beyond AI itself, though these are also beyond our scope here.

## Recommendations

This report highlights the challenges to sustaining an effective, competitive workforce as AI becomes pervasive in the workplace. AI-driven disruption will place new demands on education, policy, and business. Given that AI-related shifts in skills and jobs are already evident, we recommend urgent action in the following areas:

- **Invest in educational research that broadens AI literacy and access**, including understanding AI's capabilities and limits. Given the pervasive impact of AI, it is critical to equip students from all disciplines with basic AI literacy to engage in the workplace. This includes a focus on the skills needed to interpret responses critically. Importantly, the integration of AI into different sectors of society necessitates a workforce with expertise beyond computing in order to situate AI's limits and impacts in appropriate contexts.

- **Expand AI education funding access to non-computing disciplines.** As AI grows in capability and access, it will be important for those trained in non-computing disciplines to learn how to apply AI to the challenges of their fields, understand the opportunities and pitfalls, and adapt their profession as AI capabilities evolve. Just as interdisciplinary *CS+X* programs were developed to blend computer science with other disciplines (*X*) to meet [interdisciplinary computational demands](), *X+AI* could be a complementary approach, where training in specific disciplines is interwoven with AI concepts relevant to that field. Not only should AI literacy be pervasive in education, but the practices and pedagogy that inform AI education should be crafted through an interdisciplinary effort to capture AI's broad impact across various industries and society. To achieve this goal, funding must extend beyond traditional computing disciplines, and initiatives should encourage coordinated educational efforts.

- **Invest in AI education research that broadens educational contexts.** Developing the skills and competencies needed to engage with AI in the workforce demands investment in educational research beyond traditional four-year universities. Expanding access to AI education through other educational pathways, such as community colleges and professional training programs, will require funding and resources to advance AI infrastructure and expertise. Additionally, expanding entry points to AI literacy for current professionals will be critical to building long-term resiliency to technological change.

- **Continue to invest in research on Human-AI teaming:** Shifting the focus from job replacement to more complementary roles of AI requires continued investment in low-barrier interfaces between humans and computing platforms and extrapolating Human-AI teaming scenarios to all domains. An area of particular importance is for AI to play a larger role in upskilling humans to new AI capabilities, enabling humans and AI to co-evolve complementary skills. In [The Silent Shift: How AI Stealthily Reshapes our Work and Future](), the authors note that in the AI-accelerated future, "adaptability becomes the paramount skill."

- **Consider grants to better support access to AI across communities with fewer resources.** Community colleges will likely need assistance in funding AI integration into educational programs. A similar need exists at primary schools, particularly in low-resource districts. Comparable support may be required for small, non-technical businesses that may lack the resources to transition their workforce and business operations to the latest AI capabilities.

- **Expand education in durable core competencies.** Several core competencies will become increasingly necessary across the workforce, including systems thinking, data-literacy, critical analysis, experimental design, collaboration, and resilience. These competencies apply broadly across professional disciplines and trade careers and are vital to equipping the workforce to be more productive as change accelerates. Developing skill in these areas is of greater long-term value than training in specific software tools or ecosystems.

- **Partner with industry in reskilling and upskilling the workforce:** Just as individuals will need to be more resilient and adaptive, large and small companies will need to thrive amid AI-accelerated change. Hiring alone will not suffice for companies to keep their workforce at the leading edge. Industry must adopt a larger, forward-looking approach to continual workforce upskilling.

- **Organizations should invest in upskilling beyond their own workforce.** Investment in education and training can be leveraged across organizations. Professional societies can play a key role in creating synergies and managing shared educational resources. The AI industry can contribute as well, but there is a need for general-purpose training that is not tied to any single proprietary framework. Some executives are already shifting from the traditional employee-centric workforce model to "[workforce ecosystems](#)." In addition to employees, workforce ecosystems include external workers such as contractors, service providers, gig workers, and even software bots. Employers should consider a workforce ecosystem approach to their AI upskilling strategy. The government should support a broad, non-proprietary approach to AI training to avoid creating barriers to entry in the fast-developing field of AI.

- **Strengthen education in Responsible Use of AI**, **in both computing and non-computing curricula, and in industry and government settings.** Broad awareness across education, industry, policy and governance is necessary, and meeting this need will require the continuous evolution of educational tools and methods as AI capabilities mature.

## Acknowledgments

We thank CRA-I Steering Committee and staff members, including Mary Hall, Fatma Ozcan, Chris Ramming, Vivek Sarkar, Divesh Srivastava, and Helen Wright, for their help in framing this paper and their inputs and feedback on recommendations.

---